\documentclass[prl,twocolumn,amsmath,amssymb,groupedaddress,superscriptaddress]{revtex4}
\usepackage[dvips]{graphicx}
\usepackage{float,graphicx}
\usepackage{epsfig}

\def\be{\begin{equation}}

\def\ee{\end{equation}}

\def\qq{{\bf q}}
\def\kk{{\bf k}}

\def\LL{\textmd{L}}
\def\aat{\textmd{at}}
\def\eeff{\textmd{eff}}
\def\be{\begin{equation}}
  \def\ee{\end{equation}}
\def\bea{\begin{eqnarray}}
  \def\eea{\end{eqnarray}}
\def\ttot{\textmd{tot}}

\begin{document}

\title{Detection of Spin Correlations in Optical Lattices by Light Scattering}

\author{In\'es de Vega}
\author{J. Ignacio Cirac}
\author{D. Porras}

\affiliation{Max-Planck-Institut f\"ur Quantenoptik,
Hans-Kopfermann-Str. 1, Garching, D-85748, Germany.}

\begin{abstract}
%{\bf We show that photons with a certain polarization which are scattered within an off-resonant Raman type of experiment, are coupled to certain atomic spin operators. Hence, correlations of arbitrary combinations of an arbitrary number of spin operators can be obtained by measuring correlations between photons with different polarizations.}
We show that spin correlations of atoms in an optical lattice can be
reconstructed by coupling the system to the light, and by
measuring correlations between the emitted photons.
This principle is the basis for a method to characterize states in
quantum computation and simulation with optical lattices. As examples, we study
the detection of spin correlations in a quantum magnetic phase, and the
characterization of cluster states.
\end{abstract}

\date{\today}

\maketitle

Ultracold atoms in optical lattices open exciting prospects
for the investigation of quantum many--body phases in a highly
controllable setup.
For example, spin--spin interactions between atoms in a
Mott insulator state can be tuned in such a way that
one can simulate a rich variety of models from quantum magnetism
\cite{DDL}.
Furthermore, this system is ideally suited to implement
a quantum register for quantum computation
\cite{DJ98}, and to realize multiparticle entangled states with cold
controlled collisions \cite{BCJD99,JBCGZ99}.
An important benchmark in this direction has been the creation of a cluster
state \cite{MGWRHB03}, since it can be used as a resource for
measurement based universal quantum computation \cite{RB01,RBB03}.

A major disadvantage of this setup is the fact that atoms are separated by optical
wavelengths, and thus it is difficult to address them individually.
This impedes one to measure directly the
spatial dependence of spin--spin correlations, which is
essential in the characterization of quantum phases.
For these reasons the development of accurate methods to measure properties of
atomic operators is basic to the usefulness of optical lattices for
quantum computation and simulation.
A possible solution relies on the detection of the atom number
distribution in time of flight (TOF) imaging.
Within this technique
atomic density operators in momentum space are measured by taking absorption images
of the expanding cloud, after having switched off the trapping potentials.
From TOF images density--density correlation functions in momentum space
\cite{ADL04,FGWMGB05,GRSJ05} are reconstructed
by detecting statistical correlations between different images.
On the other hand, one may use quantum non-demolition techniques on the atom--light interaction in order to determine certain atomic collective observables \cite{EZS07,ERRLPA07}.

%In principle, TOF allows then the reconstruction of
%the diagonal terms of the one-particle correlation
%function. Nevertheless, it has recently been shown \cite{D06} that
%application of two consecutive impulsive Raman pulses at the beginning
%of the expansion, allows the measurement of non-diagonal terms in the momentum space.

In this letter we describe an alternative
approach which allows one to measure spin correlations
without switching off the optical potential. 
Furthermore, it is not necessary to drive the atoms out of the strongly correlated regime, thus avoiding certain complications that are met in TOF measurements of interacting fermions.
%Furthermore, atoms do not have to be driven out of the strongly correlated regime previous to the measurement, as in TOF experiments.
Our method is based on off--resonant scattering of an incident
laser with the trapped atoms.
Scattered photons with a certain polarization are coupled to certain spin operators. In that 
%Scattered photons carry information on atomic spin operators, in such
way correlations between photons emitted in different
directions are proportional to ground state correlations of spin operators in
momentum space. Different detection schemes, such as photon counting or homodyne
detection, may give information of different types of correlations.
One of the strengths of the method is that it is based on photon
detection, and not on atom detection. For this reason, detection of any
possible spin--spin correlation is achieved here naturally, just by 
controlling the polarization of the lasers and of the detected photons. The correlation of an arbitrary number of spin operators can be achieved by considering correlations of different detections.

%(discuss....does this mean that we can measure many photons scattered by the same atom?)
%\textit{Formalism.}
We consider an optical lattice filled with $N$ atoms in a Mott
insulator state with at most one atom per site.
Each atom carries an arbitrary ground state hyperfine
spin $J$, and has an optical dipole transition of
frequency $\omega_0$ to an excited states manifold of spin $J'$.
This transition is coupled with an off--resonant laser of frequency
$\omega_L$, wavevector ${\bf k}_L$, and a certain linear polarization
$\sigma_L$.
Photons are then scattered with momentum ${\bf k}$ and two different polarizations
$\beta$, such that $\beta \neq \sigma_L$. 
Light modes with momentum $\kk$ are coupled to atomic operators
$J^{\Delta \kk}_\beta
= 1/\sqrt{N} \sum_j J_\beta^j e^{i {\Delta \kk} \cdot{\bf r}_j }$, with
$\Delta{\bf k}={\bf k}-{\bf k}_L$,  where
$J_\beta^j$ are atomic spin operators, and
the vector ${\bf r}_j$ denotes the position  of the atoms in the
lattice. This allows us to describe correlations of spin operators with different momentum, which as it will be later discussed provides us in principle with full information to reconstruct spin correlations in the position space.
%With at most one atom per site, and a small density of holes,
%the index ${\bf r}_j = (n,m,l)r_0$, with $r_0$ the lattice distance.
We present the following results:
(i) By measuring correlations between photons with momenta $\kk$ we
reconstruct correlations between atomic spin operators
$J^{\Delta{\bf k}}_\beta$, such that $k = k_\LL$.
Full access to correlations both in momentum and position space requires that
$k_L r_0 > \sqrt{2} \pi$ ($\sqrt{3} \pi$) in $2D$ ($3D$)
square lattices, where $r_0$ is the lattice constant. However, these conditions are not always necessary to characterize a state in the momentum space.
(ii) Measured correlations are in the ground state provided that the
measurement time $T \ll 1/\Gamma$, where $\Gamma$ is the spontaneous
emission rate.
The validity of this approximation is studied by
simulating the emission properties of interacting atoms in a lattice by
means of a bosonic description of spin excitations. 
%This approach,
%up to our knowledge, is presented here for the first time.
(iii) As an illustration of applications,
we discuss the measurement of a magnetic phase, an example
that is relevant to quantum simulation, as well as a measurement of a
cluster state in a $2D$ lattice, which has an important application in quantum computation.

The incident laser has a large detuning $\Delta = \omega_{J'}-\omega_L$,
with respect to the uppermost level of $J'$, with frequency
$\omega_{J'}$.
In that situation, the excited manifold can be adiabatically
eliminated \cite{K05,J03},
and the evolution of any operator $A$
acting on the spin $J$ manifold,
can be written in terms of an effective Hamiltonian for light--matter interaction
$(d/dt)A = i/\hbar [H^{\eeff}_\textmd{lm},A]$.
Along the same lines as in \cite{K05,J03,EZS07},
the effective Hamiltonian can be written as
\begin{equation}
H^{\eeff}_{\textmd{lm}} = i g \sum_{j}
\left({\bf E}_{\ttot}^{j \dagger} \cdot  ( {\bf J}_j \times {\bf
    E}_{\ttot}^j ) + h.c.\right),
\end{equation}
where $g=d^2 a_1 /\Delta$, $d$ is the atomic dipole matrix element,
and $a_1$ is a constant that depends on the particular transition considered.
The electromagnetic field is usually expressed in terms of the two orthonormal
polarization vectors $\hat{\bf e}_{{\bf k} l} \bot \kk$
($l = 1, 2$) and the corresponding creation (annihilation)
operators $a^\dagger_{{\bf k}l}$ ($a_{{\bf k}l}$).
Nevertheless, it is more convenient for us to express
${\bf E}_{\ttot}^j$ in the laboratory frame.
We decompose the field in the following way:
 ${\bf E}_{\ttot}^j = \sum_\sigma(\widetilde{{\bf E }}_{\sigma}^{j -} +
 \widehat{{\bf E}}_{\sigma}^{j +} +{\bf E}^{ j +}_L )$,
where $\widetilde{{\bf E}}_{\sigma}^{j +}=\sum_{{\bf k}} \epsilon_{{\bf
k}} a_{{\bf
k}\sigma}e^{-i\Delta^+_{k}t+i{\bf k}\cdot{\bf r}_j} \hat{\bf \sigma}$,
and
$\widehat{{\bf E}}_{\sigma}^{j
+} =\sum_{{\bf k}} \epsilon_{{\bf k}}a_{{\bf k}\sigma}e^{-i\Delta^-_{k}t+i{\bf
k}\cdot{\bf r}_j}\hat{\bf \sigma}$ are fast and slowly rotating terms
respectively \cite{rot}, and ${\bf E}^{j+}_{ L} = \epsilon_L e^{-i
  {\bf k}_L \cdot {\bf r}_j}{\hat{\bf \sigma}}_L $ is the laser field.
The constants are $\epsilon_k = \sqrt{\hbar\omega_k /(2\epsilon_0
  \upsilon)}$, $\Delta^\pm_{k}=\omega_k \pm \omega_L$, and
${\sigma}_L,\sigma=x,y,z$.
In addition, we have defined photon operators in the laboratory frame,
$a_{{\bf k}x}=\cos\theta \cos\phi a_{{\bf k}1}
+\sin\phi a_{{\bf k}2}$, $a_{{\bf k}y}=\cos\theta \sin\phi a_{{\bf
k}1} -\cos\phi a_{{\bf k}2}$ and $a_{{\bf k}z}=-\sin\theta a_{{\bf
k}1}$, with $\theta$ and $\phi$ the angular coordinates of the wave
vector ${\bf k}$ in the laboratory frame.

Let us consider as an example the $H^{\eeff}$ when the laser
polarization is $\sigma_L = z$,
\begin{eqnarray}
%&&H^{eff}=i\Gamma_0 \sum_j \epsilon_L \left[J^j_x \left((E^{j-}_y
%-\check{E}^{j-}_y)-(E^{j+}_y
%-\check{E}^{j+}_y)\right)\right.\nonumber
%\\&&
%\left.- J^j_y \left( (E^{j-}_x -\check{E}^{j-}_x)-(E^{j+}_x
%-\check{E}^{j+}_x)\right)\right],
H^{\eeff}=ig \sum_j \left[J^j_x (E^{j-}_y-E^{j+}_y)-J^j_y (E^{j-}_x-E^{j+}_x)\right]\label{Hamiltonian2}
\end{eqnarray}
where
$E^{j-}_\sigma=\widetilde{E}^{j-}_{\sigma}e^{-i{\bf k}_L\cdot{\bf r}_j}-\widehat{E}^{j-}_{\sigma}e^{i{\bf k}_L\cdot{\bf r}_j}$, for $\sigma=x,y$.
% $E^{j-}_\beta=\sum_{{\bf
%k}}\epsilon_{{\bf k}}\hat{e}_{{\bf k}\beta} a^{\dagger}_{{\bf
%k}\beta}e^{i\Delta^-_{k}t-i{\bf k}\cdot{\bf r}_j}$, and
%$\check{E}^{j-}_\beta=\sum_{{\bf k}}\epsilon_{{\bf
%k}}\hat{e}_{{\bf k}\beta}a^{\dagger}_{{\bf
%k}\beta}e^{i\Delta^+_{k}t-i{\bf k}\cdot{\bf r}_j}$, with
%$\beta=x,y$.
The Hamiltonian (\ref{Hamiltonian2}) describes a coupling between the
emitted $y$-polarized ($x$-polariced) photons and the spin operators
$J^j_x$ ($J^j_{y}$).
Different laser polarizations give rise to the scattering of photons
that are coupled to other spin operators.
%For instance, when $\sigma_L=y$, photons are emitted with
%$\sigma=x,z$ polarizations,
%and they are coupled to $J^j_z$ and $J^j_x$ respectively.
%As we will see later, this particular set up can be used to detect cluster states.

We show first how the light--matter coupling (\ref{Hamiltonian2})
allows us to measure equal--time spin correlations of the kind
$\langle J^{\Delta{\bf k}}_\alpha J^{-\Delta{\bf k}}_\beta\rangle$,
which are very useful in the characterization of many--body spin phases.
As we show below they are related to
$N^{{\bf \hat{ k}}}_{\alpha\beta}(T)=v / (2\pi)^3 \int dk k^2 \langle
a^\dagger_{{\bf k}\alpha}(T)a_{{\bf k}\beta}(T)\rangle$, where $v$
is the quantization volume.
The diagonal elements of this quantity are the number of photons
emitted in the direction $\hat{\kk}$ during a time $T$, whereas
nondiagonal terms may be readily obtained by rotating the polarizations of the
photons prior to the measurement.
From the Heisenberg equations of the field operators, we find
\begin{eqnarray}
&&N^{{\bf \hat{ k}}}_{\alpha\beta}(T)= N\left(\frac{L}{2\pi}\right)^3 \int dk k^2 (g
\epsilon_k \epsilon_L)^2
\sum_{\alpha'\beta'}\mathcal{M}_{\alpha\alpha'}^{\bf k}\mathcal{M}_{\beta\beta'}^{\bf
k}\nonumber\\ &&\int^T_{-T} d\tau \int_0^T dt e^{-i\Delta_{\bf
k}\tau} \langle J^{\Delta{\bf
k}}_{\alpha'}(t+\tau/2)J^{-\Delta{\bf
k}}_{\beta'}(t-\tau/2)\rangle, \label{counting}
\end{eqnarray}
where the sum goes over $\alpha',\beta'=x,y,z$, and we have defined
$\mathcal{M}^{\bf k}_{\alpha\alpha'}=\sum_\beta \epsilon^{\sigma_L
  \alpha\alpha'}[a_{{\bf k}\alpha'},a_{{\bf k}\beta}^\dagger]$, where
$\epsilon^{\sigma_L \alpha\alpha'}$ is the Levi--Civita symbol.
A few approximations can be considered in order to simplify the expression (\ref{counting}).
First, the evolution of the atomic correlation due to the atomic Hamiltonian
$H_{\aat}$ occurs in a time scale
$T_A = 1/\epsilon$
(where $\epsilon$ is a typical eigenenergy of $H_{\aat}$),
that is much larger than
the light-matter interaction processes that are here considered, and can therefore be neglected.
Second, $T$ can be made short enough as to ensure that the dependency
of the correlation over $\tau$ can be neglected.
Finally, the condition $T\gg\tau_C$ (where $\tau_C$ is the
environmental decaying time),
allows us to extend to infinity the integration limits of the first integral. Hence, we can write
\begin{eqnarray}
N^{{\bf \hat{ k}}}_{\alpha\beta}(T)\approx N \Gamma_0
\sum_{\alpha'\beta'}\mathcal{M}_{\alpha\alpha'}^{\bf k}\mathcal{M}_{\beta\beta'}^{\bf
k}\int_0^T dt \langle J^{\Delta{\bf k}}_{\alpha'}(t)J^{-\Delta{\bf
k}}_{\beta'}(t)\rangle, \label{counting2}
\end{eqnarray}
where the constant
$\Gamma_0=g^2 \left(\frac{L}{2\pi}\right)^3 \pi \epsilon_L
\int dk k^2 \epsilon^2_k \delta(k-k_L)$ is the spontaneous
emission rate of a single atom.
Moreover, if
$T \ll 1/\Gamma$, where $1/\Gamma$ is an estimate of the spontaneous emission decaying time of the system that may be renormalized by collective effects, we can write
\begin{equation}
\tilde{N}^{{\bf \hat{k}}}_{\alpha\beta}(T)\approx
N \Gamma_0 \sum_{\alpha'\beta'}\mathcal{M}_{\alpha\alpha'}^{\bf k}
\mathcal{M}_{\beta\beta'}^{\bf k}T
\langle J^{\Delta{\bf k}}_{\alpha'}(0)J^{\Delta{\bf
    k}}_{\beta'}(0)\rangle.
\label{N.tilde.definition}
\end{equation}
The validity of this approximation will be later studied in more detail,
since the measuring time $T$ has to be also long enough as to ensure
that a sufficient number of photons to characterize the state is detected.

Apart from the photon number, other operators like the field
quadratures are linked to spin observables.
Let us consider detection in the far field limit, so that the detector
is placed at a position {\bf R} such that $k_L R\gg
1$. In that case, it can be
shown that the modes that contribute more to the field are those
with wavevector in directions ${\bf\hat{k}} = {\bf \hat{R}}$, with
${\bf \hat{R}}$ a unit vector in the direction of the detector
\cite{L70}.
In addition, due to energy conservation, the largest contribution to the
emitted field comes from $k = k_L$.
Hence, the positive component of the emitted field corresponding to
the $\alpha$ polarization ca be written as
$E^{+}_\alpha({\bf r},T) \approx
\sqrt{N}\frac{\Gamma_0}{g} e^{-i\omega_L t+i k_L R}
\sum_\beta M^{\bf \hat{R}}_{\beta\alpha} J^{-\Delta {\bf R}}_\beta (0)$,
where $\Delta{\bf R}=k_L {\bf \hat{R}}-{\bf k}_L$.
Here we have followed similar approximations as in the derivation of
(\ref{counting2}),
and also that $J^{\Delta{\bf k}}_\beta (\tau)\approx J^{\Delta{\bf k}}_\beta (0)$ for
$\tau\epsilon[0,T]$, provided again that $T \ll 1/\Gamma$.
By performing homodyne detection one may, for example, measure
the quadrature
${\mathcal X}_\alpha^{\bf k}=\epsilon_{\bf k} (a^\dagger_{{\bf
    k}\alpha}+a_{{\bf k}\alpha})$, which is related to the
spin--operators in the following way:
\begin{eqnarray}
{\mathcal X}_\alpha^{\bf k}(T)=
%{\mathcal X}_\alpha^{{\bf k}}(0)+
2 \sqrt{N}\frac{\Gamma_0}{g}
\sum_{\beta}\mathcal{M}^{\bf k}_{\alpha\beta}J^{\Delta{\bf
k}}_{C \beta} (0), \label{quadratures}
\end{eqnarray}
where ${\bf k}=k_L{\bf \hat{R}}$, and $\beta=x,y,z$.
We have defined
$J^{\Delta{\bf k}}_{C \beta} (0)
= (J^{\Delta{\bf k}}_\beta (0)+J^{-\Delta{\bf k}}_\beta(0))/2
= \sum_j \cos(\Delta{\bf k}\cdot{\bf r}_j)J^j_\beta /\sqrt{N}$.
From (\ref{quadratures}), we see that a certain quadrature is
proportional to a combination of system spin operators.
If we want to detect only a certain spin component, for instance $J^{\Delta{\bf k}}_{C x}$,
we should fix the detector in $\theta$ and $\phi$ such that only
$\mathcal{M}_{\alpha x}\neq 0$.
Different values of $\Delta{\bf k}$ are scanned by changing
${\bf \hat{k}}_L$, what would provide us with information of atomic
spins (and their correlations) within a $2D$ slice in momentum space.
Spin operators in the whole $3D$ momentum space, may be obtained by considering
measurements in which both the detector and the laser are moved.
Note that to extract information on spin operators from optical
measurements requires to invert the matrix
${\bf \mathcal{M}}^{\bf k}´=\{\mathcal{M}^{\bf k}_{\alpha,\beta}\}$,
what in general can be done except for some particular values of
$\theta$ and $\phi$ such that $det[{\bf \mathcal{M}}]\neq 0$.

%\textit{Applications}
We now turn to study further the approximation
$N_{\alpha,\beta}(T) \approx \tilde{N}_{\alpha,\beta}(T)$, which
allows us to reconstruct ground state correlations from the emitted
photons.
The idea is to calculate how much of the information we get from the atoms corresponds to their ground state.
%Our goal is to calculate the relative error,
%\begin{equation}
%(E^\alpha_R)^2 =
%\int d\Omega_{\bf k}(N_{\alpha\alpha}^{{\bf k}}-\tilde{N}_{\alpha\alpha}^{{\bf
%k}})^2 / \int d\Omega_{\bf k}(N_{\alpha\alpha}^{{\bf
%k}})^2.
%\label{def.ER}
%\end{equation}
The study of the radiative emission of an
ensemble of interacting atoms poses a complicated many--body
problem.
We address it here by considering that atoms have a well
defined magnetic ordering along the $z$ axis, such that
$[ J^j_x, J^l_y ] = i \delta_{j,l}$, in a Holstein--Primakoff (HP)
approximation.
Our detection method relies on $E^\alpha_R$ being small up to $T$ such
that a sufficient number of photons has been emitted to ensure a good
detection efficiency.
Within the HP approximation, Hamiltonian (\ref{Hamiltonian2}) yields
the following system of evolution equations:
\begin{eqnarray}
\frac{\partial \langle J^{\bf q}_x \rangle}{\partial t}
&=&
-U\int
d\Omega_{\bf k} \delta_{1/L}^{{\bf q}-\Delta{\bf k}} \left[M^{\bf
k}_{xy} \langle J^{\Delta{\bf k}}_{y}\rangle +M^{\bf k}_{xx}
\langle J^{\Delta{\bf
k}}_{x}\rangle \right],
\nonumber\\
\frac{\partial \langle J^{\bf q}_y \rangle}{\partial t}
&=&
U\int
d\Omega_{\bf k} \delta_{1/L}^{{\bf q}+\Delta{\bf k}} \left[M^{\bf
k}_{yy} \langle J^{-\Delta{\bf k}}_{y}\rangle +M^{\bf k}_{yx}
\langle J^{-\Delta{\bf k}}_{x}\rangle \right],\label{system}
\label{operator.evolution}
\end{eqnarray}
where $U = 4 g^2 \Gamma_0$,
and the spin dynamics induced by
$H_{\textmd{at}}$ during the radiative time scale has been neglected.
We have also assumed that atoms are distributed in a lattice with
single occupation, something that does not alter the qualitative
features of the emission process.
$\delta_{1/L}^{{\bf q}} = \sum_j e^{i{\bf q}\cdot{\bf r}_j}$
is a function centered at ${\bf q}=0$ with width $1/L$.
In the limit $1/L \ll d_0$, spin operators vary much more slowly than
$\delta_{1/L}^{\qq}$ in momentum space, such that
one can approximate
$\int d\Omega_{\bf k} {\cal M}^\kk_{\alpha \beta}
\delta_{1/L}^{{\bf q} \mp \Delta{\bf k}} \langle
J^{\pm\Delta{\bf k}}_{\beta} \rangle
\approx
\langle J^{\pm{\bf q}}_\beta\rangle
\int d\Omega_{\bf k} {\cal M}^\kk_{\alpha \beta} \delta_{1/L}^{{\bf q} \mp \Delta{\bf k}} $,
and  (\ref{system}) becomes a closed
system for each atomic operator with momentum ${\bf q}$.
Finally, from the  quantum regression theorem, (\ref{system}) can be
used to evolve two operator averages, and calculate the emission
pattern by means of Eq. (\ref{counting2}).

To present a definite situation, we consider atoms with spin
$1/2$, with
$H_{\aat} = B \sum_j J^j_z + J \sum_{<ij>} J_x^i J_x^j$,
where $B$ is an external magnetic field, we consider a ferromagnetic
interaction ($J < 0$),  and $<ij>$ refers to nearest neighbors.
In the regime $B > |J|$, and far from the critical point, the system is
in a paramagnetic phase with spins aligned along the $\hat{z}$ axis.
The HP approximation yields an adequate description of the quantum
fluctuations in the ground state. Within this approximation the
problem can be rewritten in momentum space,
$H_{\aat} =
\sum_{\bf q}(2 B + J_{\bf q}) J^{\bf q}_x J^{-{\bf q}}_x + J^{\bf q}_y
J^{-{\bf q}}_y$, where we have used the Fourier transformed spin--spin
interaction,
$J_{\bf q} = 2 J(\cos(q_x r_0) + \cos(q_y r_0) +
\cos(q_z r_0))$.
Note that the description of the many--body problem in terms of spin--waves
suits perfectly our purpose, since it also allows us to describe the
emission process.
The ground state correlations within this approximation are
$\langle J^{\bf q}_y J^{-{\bf q}}_y \rangle = \sqrt{1 + J_{\bf q}/4B}$,
$\langle J^{\bf q}_x J^{-{\bf q}}_x \rangle = 1/2 \sqrt{1 + J_{\bf q}/4B}$,
$\langle J^{\bf q}_x J^{-{\bf q}}_y \rangle = i/2$,
and $\langle J^{\bf q}_y J^{-{\bf q}}_x \rangle = - i/2$.
They are
used as initial condition to the set of equations obtained by means of
the evolution equations (\ref{operator.evolution}).

\begin{figure}
  \resizebox{\linewidth}{!}{%
    \includegraphics{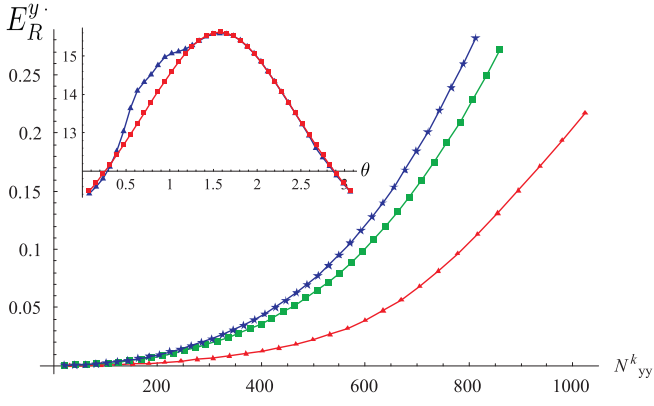}}
  \caption{Relative error between ${\cal N}^{\Omega_{\theta,\phi=0}}_{yy}$ and $\tilde{N}^{\Omega_{\theta,\phi=0}}_{yy}(T)$, with respect to the number of emitted photons with
     $\alpha=y$, integrating all over $\theta$ for $\phi=0$. Triangles: $J/B=-0.5$, Boxes: $J/B=-0.1$, Stars: $J/B=-0.01$.; Inset: Number of $y$-polarized photons for each value of $\theta$ with $\phi=0$, with $T=0.001$ (units of $\Gamma_0=1$), and $J/B=-0.5$. Triangles: ${\cal N}^{\Omega_{\theta,\phi=0}}_{yy}=\Gamma_0 N \int_0^T dt \langle J^{\theta,\phi=0}_x (t)J^{\theta,\phi=0}_x (t)\rangle $ (Emitted photons). Squares: $\tilde{N}^{\Omega_{\theta,\phi=0}}_{yy}(T)= \Gamma_0 N T \langle J^{\theta,\phi=0}_x (0)J^{\theta,\phi=0}_x (0)\rangle$ (Emitted photons that correspond to ground state)}
\label{f1}
\end{figure}

As an example of measurement,
we show in Fig. (\ref{f1}) the quantity ${\cal N}^{\Omega_{\theta,\phi=0}}_{yy}$,
for a certain small $T$ such that most of the photons come from the system ground state.
In addition, the relative error, 
\begin{equation}
(E^y_R)^2 =
\int d\cos(\theta)({\cal N}^{\Omega_{\theta,\phi}}_{yy}-\tilde{N}^{\Omega_{\theta,\phi}}_{yy})^2 / \int d\cos(\theta)({\cal N}^{\Omega_{\theta,\phi}}_{yy})^2,
\label{def.ER}
\end{equation}
%between ${\cal N}^{\Omega_{\theta,\phi=0}}_{yy}$
%and $\tilde{N}^{\Omega_{\theta,\phi=0}}_{yy}$
%(which tells us how much of the information we get from
%the atoms corresponds to their ground state) 
where we have fixed $\phi=0$, is plotted in Fig. (\ref{f1}) with respect to the number of emitted photons
(proportional to $T$).
It can be seen that the error remains relatively small
when enough number of photons (of the order of $600$)
have been emitted.
% within the solid angle.
% along the solid angle $\Omega_{\theta,\phi=0}$.
We stress that the radiative emission from an interacting spin system is
indeed an interesting problem by its own, which shows the interplay between
quantum dissipation and many--body effects.

The information gathered by measuring in different directions,
may  be used to reconstruct atomic correlations in the position space.
To study the conditions which are required for this task, we focus on
%the particular example of a 3D optical lattice in which atoms are
%correlated in the 2D $x$--$y$ planes, and uncorrelated in the $z$
%direction.
the particular example of a $2D$ optical lattice.
This case is relevant to the characterization of cluster
states and experiments which simulate high--Tc superconductors.
Consider also for concreteness that the incident laser propagates
along the $z$ direction, and that the lattice is defined within the $x$-$y$ plane.
Following (\ref{quadratures}), an homodyne detection of the quadrature $\mathcal{X}^{\bf k}_{\alpha}$ allows one to measure the operator $J_{C \beta}^{\Delta{\bf k}}$,
%where
%\begin{equation}
%J_{C \beta}^{\Delta{\bf k}} = \frac{1}{\sqrt{N}}
%\sum_{m,n,l} \cos{( \tilde{k}_x n \pi + \tilde{k}_y m \pi)}
%J^{(n,m)}_{\beta},
%\label{Jc}
%\end{equation}
where now we have that $\Delta{\bf k} = k_\LL  (\sin\theta \cos\phi, \sin\theta \sin\phi, \cos\theta - 1)$.
%Due to the energy conservation there are only two
%independent variables, the angles $\theta $ and $\phi$,
%so that we can write
%$\tilde{k}_z = \frac{r_0 k_L}{\pi}
%\left(\sqrt{1 - (\frac{\pi}{r_0 k_L})^2(\tilde{k}_x^2 +
%\tilde{k}_y^2)} - 1 \right)$.
%Correlations of spin operators in space are obtained as correlations of the inverse Fourier transform of (\ref{Jc}), $J^{(n,m,0)}_{\beta}$. Note that here we have specified the atomic index $j$ in terms of the lattice site indexes of a two dimensional lattice $(n,m,0)$.
Then the spin operator $J^{(n,m)}_{C\beta}=J_{\beta}^{(n,m)}+J_{\beta}^{(-n,-m)}$ in the position space, with $(n,m)$ denoting a particular site within the lattice, can be obtained as an inverse Fourier transform of $J_{C \beta}^{\Delta{\bf k}}$,
%The position index of an atomic operator within a
%$2D$ lattice can just be specified by $(n,m)$.
%The inverse Fourier transform can be then written as
\begin{eqnarray}
J^{(n, m)}_{C\beta}=C\int^1_{-1}d\tilde{k}_x\int^1_{-1}d\tilde{k}_y \cos(\tilde{k}_x n
\pi+\tilde{k}_y m \pi)J_{C\beta}^{\Delta{\bf k}},
%{\mathcal{F}_{k_x,k_y,L}}
%(\tilde{k}_x,\tilde{k}_y,L)},
\label{limits}
\end{eqnarray}
where we have defined $C=\sqrt{N}/2$, and the integration variables as $\tilde{\bf k}=(r_0/\pi)\Delta{\bf k}$.
%and
%$\mathcal{F}_{k_x,k_y,L}=\csc(\frac{\pi k_z}{2})\sin(\frac{\pi (L+1)k_z}{2})$.
By changing the detection angles
($\theta,\phi$),
it is possible to measure the quantity
$J_{C\beta}^{\Delta{\bf k}}$
for values of
${\Delta{\bf k}}$
that lie within a circle of radius
$R_0 = 2 k_\LL r_0 / \pi$.
The integral (\ref{limits}) has to be sampled with a set of
values of
$J_{C\beta}^{\Delta{\bf k}}$
within the region of integration defined by the square
$\tilde{k}_x \epsilon[-1,1]$, and
$\tilde{k}_y \epsilon[-1,1]$.
Therefore, a basic requirement to obtain $J^{(n, m)}_{C\beta}$ is to chose
$r_0 k_L \geq \sqrt{2}\pi$, so that the integration region is
contained within the circle $R_0$. Note that this implies that $k_L/k_{latt}\ge\sqrt{2}$, where $k_{latt}$ is wavelength of the standing wave lasers.
An additional homodyne detection of the quadrature
$\mathcal{P}_\alpha^{\bf k}=i\epsilon_{\bf k}(a^\dagger_{{\bf
    k}\alpha}-a_{{\bf k}\alpha})$, allows one to obtain the quantity
$J^{(n,m)}_{S\beta}=J_{\beta}^{(n,m)}-J_{\beta}^{(-n,-m)}$, which
combined with (\ref{limits}) can be used to measure the operator
$J_{\beta}^{(n,m)}$. Correlations in 3D can be obtained provided that both the direction of photon detection, and the
direction of the incident laser are tuned to scan the whole
momentum space. Following the same argument as before, the condition to get spatial information 
%readily extended to correlations in 3D. In that case the required condition is 
%In such a case, operators in space can be recovered if the condition
%can be recovered if the condition 
is $r_0 k_L \geq \sqrt{3} \pi$. 
% is fulfilled.

Once the operators $J_{\beta}^{(n,m)}$ are obtained as
the inverse Fourier transforms of field quadratures,
it is straightforward to calculate their correlations.
These spatial correlations are useful to characterize some interesting
states, like for instance cluster states $|\phi\rangle_{cl}$.
The later are defined by a set of eigenvalue equations for the
operator $K^j = \sigma_x^j \bigotimes_{l \epsilon neigh(j)} \sigma_z^l$,
such that $K^j | \phi\rangle_{cl}=\pm | \phi\rangle_{cl}$
\cite{RB01,RBB03},
where $neigh(j)$ specifies the sites of all atoms $l$ that interact
with an atom $j$.
With this definition at hand,
a cluster state in a $2D$ lattice can be characterized by checking
that the spatial correlation
$\langle
J^{(n_1,m_1)}_{z}J^{(n_1,m_2)}_{z}J^{(n_3,m_3)}_{x}J^{(n_4,m_4)}_{z}
J^{(n_5,m_5)}_{z}\rangle$
is equal to $\pm 1$ when the operators $J_z$ are next neighbors of
$J_x$. These spatial correlations can be obtained by making an inverse
Fourier transform of the quantity
$\langle J_{z}^{\Delta{\bf k}_1}J_{z}^{\Delta{\bf k}_2}J_{
  x}^{\Delta{\bf k}_3}J_{z}^{\Delta{\bf k}_4}J_{z}^{\Delta{\bf k}_5}\rangle$,
that may be measured by considering a $y$ polarized laser and
correlations of different homodyne detections of quadratures.
Then, the inverse transform can be made following the basic relation (\ref{limits}), and the analogous relation that exists for $J^{(n,m)}_{S\beta}$.
This scheme can be readily extended to measure many operator averages
which characterize magnetic quantum phases, like for example, the valence bond
strength, $\langle {\bf J}_j \cdot {\bf J}_l \rangle$, where $j$, $l$
are nearest neighbors.

One of the main limitations of the method is the fact that we need a
good detection efficiency.
The reason is that only a few photons are emitted at each solid angle
(see Fig. (\ref{f1})), because the measurement time $T$ has to be
small enough to ensure that only a few atoms produce scattering
and the measured state is preserved. Also, just as in TOF imaging,
the reconstruction of atomic correlation functions,
requires to perform each measurement $M$ times, with $M$ large so that
the quantity $1/\sqrt{M}$ that characterizes the
statistical error for the measurement is small.
Nevertheless, one important difference with respect to TOF is
that there are no limitations regarding the laser shot noise.
In TOF experiments, the atomic noise has to exceed the
shot noise of the prove laser. Here, the fluctuations of the prove
laser can be easily distinguished and eliminated,
since the laser polarization is different
from the polarization of the emitted photons that we measure.
Since our model is valid for measuring atoms in a Mott state,
a further source of error would be the appearance of
a superfluid region in the borders of the atomic cloud.
This problem can be overcome by focusing the
scattering laser to the center of the ensemble.

In summary, we have shown how to detect
spin correlations of an atom lattice within a Mott insulator state
without switching off the potential.
The detection scheme is based on the fact that spin correlations
in the momentum space are proportional to correlations of the photons
that are emitted in an off-resonant scattering process. 
%Provided that the measurement time is small enough, the detected spin correlations are basically in the ground state, as it has been studied by simulating the atom--light interaction by means of a bosonic type of description of the atomic observables.
Using different photon detection techniques allows to measure
different types of spin correlations that are useful to characterize
certain many body states, like magnetic phases.
A complete sampling of these correlations in the momentum space
can be used to obtain spatial correlations,
which are useful to characterize some other phases like cluster states. 

We would like to thank Miguel Aguado for helpful discussions.
Work supported by EU projects (SCALA and CONQUEST), and Cluster of Excelence Munich-Centre for Advanced Photonics (MAP).
I.D.V acknowledges support from Ministerio de Educaci\'on y Ciencia.

\begin{thebibliography}{99}
\bibitem{rot} The fast rotating term shall be included here to describe correctly the dipole-dipole interactions \cite{L70}.
\end{thebibliography}
\end{document}